\documentclass[aps,epsf,twocolumn,showpacs]{revtex4}
\usepackage{amsmath}
\usepackage{epsfig}

\begin{document}

\title{Multicritical Points and Crossover Mediating the Strong Violation of Universality: Wang-Landau
Determinations in the Random-Bond $d=2$ Blume-Capel model}

\author{A. Malakis$^1$}

\author{A. Nihat Berker$^{2,3}$}

\author{I. A. Hadjiagapiou$^1$}

\author{N. G. Fytas$^1$}

\author{T. Papakonstantinou$^1$}

\affiliation{$^1$Department of Physics, Section of Solid State
Physics, University of Athens, Panepistimiopolis, GR 15784
Zografos, Athens, Greece}

\affiliation{$^2$Faculty of Engineering and Natural Sciences,
Sabanc$\imath$ University, Orhanl$\imath$, Tuzla 34956, Istanbul,
Turkey}

\affiliation{$^3$Department of Physics, Massachusetts Institute of
Technology, Cambridge, Massachusetts 02139, U.S.A.}

\date{\today}

\begin{abstract}
The effects of bond randomness on the phase diagram and critical
behavior of the square lattice ferromagnetic Blume-Capel model are
discussed. The system is studied in both the pure and disordered
versions by the same efficient two-stage Wang-Landau method for
many values of the crystal field, restricted here in the
second-order phase transition regime of the pure model. For the
random-bond version several disorder strengths are considered. We
present phase diagram points of both pure and random versions and
for a particular disorder strength we locate the emergence of the
enhancement of ferromagnetic order observed in an earlier study in
the ex-first-order regime. The critical properties of the pure
model are contrasted and compared to those of the random model.
Accepting, for the weak random version, the assumption of the
double logarithmic scenario for the specific heat we attempt to
estimate the range of universality between the pure and
random-bond models. The behavior of the strong disorder regime is
also discussed and a rather complex and yet not fully understood
behavior is observed. It is pointed out that this complexity is
related to the ground-state structure of the random-bond version.
\end{abstract}

 \maketitle

\section{Introduction}
\label{sec:1}

The effect of quenched randomness on the equilibrium and dynamic
properties of macroscopic systems is a subject of great
theoretical and practical interest. It is well known that quenched
bond randomness may produce drastic changes on phase transitions
depending on the type of the
transition~\cite{harris74,berker90,aizenman89,hui89,berker93,falicov96}.
Thus, symmetry-breaking first-order transitions are converted to
second-order phase transitions by infinitesimal bond randomness
for spatial dimensionality $d=2$~\cite{aizenman89,hui89} and by
bond randomness beyond a threshold strength in $d>2$~\cite{hui89},
as indicated by general arguments~\cite{berker93} and in some
cases by rigorous mathematical work~\cite{aizenman89}. In
particular, this rounding effect of first-order transitions has
now been rigorously established in a unified way in low dimensions
($d\leq 2$) including a large variety of types of randomness in
classical and quantum spin systems~\cite{greenblatt09}.

Historically, the effects of disorder on phase transitions have
been studied in two extreme cases, i.e. in the limits of weak and
strong (near the percolation point) disorder. The first important
conjecture, known today as the Harris criterion~\cite{harris74},
relates the value of the specific heat exponent $\alpha$ in a
continuous transition with the expected effects of uncorrelated
weak disorder in ferromagnets. According to the Harris criterion,
for continuous phase transitions with a negative exponent
$\alpha$, the introduction of weak randomness is expected to be an
irrelevant field and the disordered system to remain in the same
universality class. On the other hand, the weakly disordered
system is expected to be in a different universality class in the
case of a pure system having a positive exponent $\alpha$. Pure
systems with a zero specific heat exponent ($\alpha=0$) are
marginal cases of Harris criterion (since the criterion does not
give any information) and their study, upon the introduction of
disorder, has been of particular interest. The paradigmatic model
of the marginal case is, of course, the general random 2d Ising
model (random-site, random-bond, and bond-diluted) and this model
has been extensively investigated and
debated~\cite{fytas08c,grinstein76,fisch78,dotsenko81,shalaev84,shankar87,ludwig87,wang90,kim94,dotsenko95,reis96,ballesteros97,selke98,mazzeo99,martins07,hasenbusch08,gordillo09}.
Several recent studies, both analytical (renormalization group and
conformal field theories) and numerical (mainly Monte Carlo (MC)
simulations) devoted to this model, have provided very strong
evidence in favor of the so-called logarithmic corrections's
scenario. According to this, the effect of infinitesimal disorder
gives rise to a marginal irrelevance of randomness and besides
logarithmic corrections, the critical exponents maintain their 2d
Ising values. In particular, the specific heat is expected to
slowly diverge with a double-logarithmic
dependence~\cite{dotsenko81,shalaev84,shankar87,ludwig87}. Here,
we should mention that there is not full agreement in the
literature and a different scenario predicts a negative specific
heat exponent $\alpha$ leading to a saturating
behavior~\cite{kim94}, with a corresponding correlation length
exponent $\nu\geq 2/d$~\cite{chayes86}.

In general, a unitary and rigorous physical description of
critical phenomena in disordered systems still lacks and
certainly, lacking such a description, the study of further models
for which there is a general agreement in the the behavior of the
corresponding pure cases is very important. Historically, such a
suitable candidate for testing the above predictions, that has
been also quite extensively studied, is the general 2d $q$-state
Potts
model~\cite{ludwig87b,chen95,wiseman95a,dotsenko95,jacobsen98,picco96}.
This model includes the Ising model ($q=2$), cases of a pure
system having continuous transitions with a positive exponent
$\alpha$ ($q=3,4$) and also the large $q$-cases ($q>4$) for which
one could observe and try to classify the above mentioned
softening of first-order transitions in 2d models. Another
similarly interesting candidate, not yet as much studied in the
random-bond version, is the 2d Blume-Capel (BC)
model~\cite{blume66,capel66}. We may note here that most of the
existing literature on the BC model with randomness concerns
randomness applied to the crystal field and/or spin glass exchange
interactions~\cite{kaufman90,puha01,salmon09,ozcelik}. As it is
well known, the pure version of the BC model undergoes an
Ising-like continuous phase transition to an ordered ferromagnetic
phase as the temperature is lowered for crystal-field couplings
less than a tricritical value and a first-order transition for
larger values of the crystal-field coupling. Therefore, this model
provides also the opportunity to study two different and very
interesting topics of the above described effects of disorder in
critical phenomena, namely the double-logarithmic scenario for the
specific heat in the regime where the 2d BC model is in the same
universality class with the 2d random-bond Ising model and also
the softening of the transition in the first-order regime.
Recently the present authors~\cite{malakis09} have considered this
model and provided strong numerical evidence clarifying two of the
above mentioned effects induced in 2d systems by bond randomness.
By implementing a two-stage Wang-Landau (WL)
approach~\cite{wang01,wang01b,malakis04,malakis05,malakis06,fytas08a,malakis09},
we presented essentially exact information on the 2d BC model
under quenched bond randomness. In this investigation, we found
dramatically different critical behaviors of the second-order
phase transitions emerging from the first- and second-order
regimes of the pure BC model and since, these second-order
transitions were found to have different critical exponents, our
study indicated an interesting strong violation of
universality~\cite{malakis09}. Namely, different sets of critical
exponents on two segments of the same critical line appeared to
describe the two regimes: still-second-order and ex-first-order.

In this paper, we extend our earlier work~\cite{malakis09}, by
implementing essentially the same two-stage WL approach
(Sec.~\ref{sec:2}) and try to give a more complete picture by
concentrating in the weak (still-second-order) regime and simulate
the model for several disorder strengths and many values of the
crystal-field coupling. The above statement means that,
effectively we will restrict our study to moderate values of the
crystal field and moderate values of the disorder, intending to
observe the frontier between the weak- and the strong-disorder
universality classes from the disappearance of the expected 2d
random Ising universality class behavior. Thus, in
Sec.~\ref{sec:3} we will produce phase diagram points for the
random-bond model but also for the pure model, reporting for the
pure case a comparison with existing estimates in the literature.
More generally, in carrying out this project we have also
considered the pure 2d BC model for several values of the
crystal-field, in the second-order regime, observing its
finite-size scaling (FSS) behavior. Sec.~\ref{sec:4} presents such
a comparative study between random and pure models concerning the
behavior of all thermodynamic parameters used in the traditional
FSS analysis of MC data. This study enables us to observe some
peculiarities of the pure model, due to the onset of
tricriticality, and compare them with the corresponding behavior
of the random model. Furthermore, we try to focus, understand and
shed light to the extent of universality of the random-bond 2d BC
model with the corresponding random-bond Ising model, for which
the scenario of logarithmic corrections seems to be the strongest
option in the current literature~\cite{gordillo09}. The prediction
of the range of such universality is far from trivial and the two
regimes (weak and strong) have many dissimilarities which are also
reflected in the ground-state structure, as further discussed
below. The attempt to estimate the range of the above mentioned
universality is accomplished by the novel idea which assumes the
truth of the double logarithmic scenario for the specific heat in
a suitable restricted range. This is presented in Sec.~\ref{sec:5}
together with some further crucial observations concerning the
behavior of the strong disorder regime, i.e. the regime where the
Ising class universality does not apply and the system has a
rather complex and yet not understood behavior. Our conclusions
are summarized in Sec.~\ref{sec:6}.

\section{Definition of the Models and the two-stage Wang-Landau approach}
\label{sec:2}

\subsection{The pure and random-bond Blume-Capel models}
\label{sec:2a}

The (pure) BC model~\cite{blume66,capel66} is defined by the
Hamiltonian
\begin{equation}
\label{eq:1}
H_{p}=-J\sum_{<ij>}s_{i}s_{j}+\Delta\sum_{i}s_{i}^{2},
\end{equation}
where the spin variables $s_{i}$ take on the values $-1, 0$, or
$+1$, $<ij>$ indicates summation over all nearest-neighbor pairs
of sites, and $J>0$ is the ferromagnetic exchange interaction. The
parameter $\Delta$ is known as the crystal-field coupling and to
fix the temperature scale we set $J=1$ and $k_B=1$. As it is well
known, this model has been analyzed, besides the original
mean-field theory~\cite{blume66,capel66}, by a variety of
approximations and numerical approaches. These include the real
space renormalization group, MC simulations, and MC
renormalization-group calculations~\cite{landau72},
$\epsilon$-expansion renormalization groups~\cite{stephen73},
high- and low-temperature series calculations~\cite{fox73}, a
phenomenological FSS analysis using a strip
geometry~\cite{nightingale82,beale86}, and, finally, a recent
two-parameter WL sampling in rather small lattices of linear sizes
$L\leq 16$~\cite{silva06}. As mentioned already in the
introduction the phase diagram of the model consists of a segment
of continuous Ising-like transitions at high temperatures and low
values of the crystal field which ends at a tricritical point,
where it is joined with a second segment of first-order
transitions between ($\Delta_{t},T_{t}$) and ($\Delta=2,T=0$).

The model given by Eq.~(\ref{eq:1}) is studied here on the square
lattice and will be referred to as the pure BC model. However, our
main focus, on the other hand, is the case with bond disorder
given by the bimodal distribution
\begin{align}
\label{eq:2}
P(J_{ij})~=~&\frac{1}{2}~[\delta(J_{ij}-J_{1})+\delta(J_{ij}-J_{2})]\;;\\
\nonumber
&\frac{J_{1}+J_{2}}{2}=1\;;\;\;J_{1}>J_{2}>0\;;\;\;r=\frac{J_{2}}{J_{1}}\;,
\end{align}
so that $r$ reflects the strength of the bond randomness. The
resulting quenched disordered (random-bond) version of the
Hamiltonian defined in Eq.~(\ref{eq:1}) reads now as
\begin{equation}
\label{eq:3} H=-\sum_{<ij>}J_{ij}s_{i}s_{j}+\Delta
\sum_{i}s_{i}^{2}.
\end{equation}

A first comparative study between the two versions (pure and
random) of the 2d BC model has been already presented in our
earlier paper~\cite{malakis09} for two values of the crystal-field
coupling corresponding to the second-order ($\Delta=1$) and
first-order ($\Delta=1.975$) regimes of the pure model where for
the random version the disorder strength $r=0.75/1.25=0.6$ was
chosen in both cases. In the next Sections our study on the 2d BC
model is extended to several values of the crystal field and
disorder (listed in the table of Sec.~\ref{sec:3}). Details of our
simulations are summarized in the next Section together with an up
to date brief sketch of our entropic scheme.

\subsection{An outline of our implementation of the Wang-Landau approach}
\label{sec:2b}

In the last few years we have used an entropic sampling
implementation of the WL algorithm~\cite{wang01,wang01b} to study
some simple~\cite{malakis04,malakis05}, but also some more complex
systems~\cite{malakis06,fytas08a,fytas08c}. One basic ingredient
of this implementation is a suitable restriction of the energy
subspace for the implementation of the WL algorithm. This was
originally termed as the critical minimum energy subspace (CrMES)
restriction~\cite{malakis04,malakis05} and it can be carried out
in many alternative ways, the simplest being that of observing the
finite-size behavior of the tails of the energy probability
density function (e-pdf) of the system~\cite{malakis05}.
Complications that may arise in random systems can be easily
accounted for by various simple modifications that take into
account possible oscillations in the e-pdf and expected
sample-to-sample fluctuations of individual disorder realizations.
In our recent papers~\cite{fytas08a,fytas08c,malakis09}, we have
presented details of various sophisticated routes for the
identification of the appropriate energy subspace $(E_{1},E_{2})$
for the entropic sampling of each random realization. In
estimating the appropriate subspace from a chosen pseudocritical
temperature one should be careful to account for the shift
behavior of other important pseudocritical temperatures and extend
the subspace appropriately from both low- and high-energy sides in
order to achieve an accurate estimation of all finite-size
anomalies. Of course, taking the union of the corresponding
subspaces, insures accuracy for the temperature region of all
studied pseudocritical temperatures.

The up to date version of our implementation uses a combination of
several stages of the WL process. First, we carry out a starting
(or preliminary) multi-range (multi-R) stage, in a very wide
energy subspace. This preliminary stage is performed up to a
certain level of the WL random walk. The WL refinement is
$G(E)\rightarrow f* G(E)$, where $G(E)$ is the density of states
(DOS) and we follow the usual modification factor adjustment
$f_{j+1}=\sqrt{f_{j}}$ and $f_{1}=e$. The preliminary stage may
consist of the levels : $j=1,\ldots,j=18$ and to improve accuracy
the process may be repeated several times. However, in repeating
the preliminary process and in order to be efficient, we use only
the levels $j=13,\ldots,18$ after the first attempt, using as
starting DOS the one obtained in the first random walk at the
level $j=12$. From our experience, this practice is almost
equivalent of simulating the same number of independent WL random
walks. Also in our recent studies we have found out that is much
more efficient and accurate to loosen up the originally applied
very strict flatness criteria~\cite{malakis04,malakis05}. Thus, a
variable flatness process starting at the first levels with a very
loose flatness criteria and assuming at the level $j=18$ the
original strict flatness criteria is now days used. After the
above described preliminary multi-R stage, in the wide energy
subspace, one can proceed in a safe identification of the
appropriate energy subspace using one or more alternatives
outlined in Refs.~\cite{malakis04,malakis05}. In random systems,
where one needs to simulate many disorder realizations, it is also
possible and advisable to avoid the identification of the
appropriate energy subspace separately for each disorder
realization by extrapolating from smaller lattices and/or by
prediction from preliminary runs on small numbers of disorder
realizations. In any case, the appropriate subspaces should be
defined with sufficient tolerances. In our implementation we use
such advance information to proceed in the next stages of the
entropic sampling.

The process continues in two further stages (two-stage process),
using now mainly high iteration levels, where the modification
factor is very close to unity and there is not any significant
violation of the detailed balance condition during the WL process.
These two stages are suitable for the accumulation of histogram
data (for instance energy-magnetization histograms), which can be
used for an accurate entropic calculation of non-thermal
thermodynamic parameters, such us the order parameter and its
susceptibility~\cite{malakis05}. In the first (high-level) stage,
we follow again a repeated several times (typically $\sim 5-10$)
multi-R WL approach, carried out now only in the restricted energy
subspace. The WL levels may be now chosen as $j=18,19,20$ and as
an appropriate starting DOS for the corresponding starting level
the average DOS of the preliminary stage at the starting level may
be used. Finally, the second (high-level) stage is applied in the
refinement WL levels $j=j_{i},\ldots,j_{i}+3$ (typically
$j_{i}=21$), where we usually test both an one-range (one-R) or a
multi-R approach with large energy intervals. In the case of the
one-R approach we have found very convenient and in most cases
more accurate to follow the Belardinelli and
Pereyra~\cite{belardinelli07} adjustment of the WL modification
factor according to the rule $\ln f\sim t^{-1}$. Finally, it
should be also noted that by applying in our scheme a separate
accumulation of histogram data in the starting multi-R stage (in
the wide energy subspace) offers the opportunity to inspect the
behavior of all basic thermodynamic functions in an also wide
temperature range and not only in the neighborhood of the
finite-size anomalies. The approximation outside the dominant
energy subspace is not of the same accuracy with that of the
restricted dominant energy subspace but is good enough for the
observation of the general behavior and provides also a route of
inspecting the degree of approximation.

The above described numerical approach was used to estimate the
properties of a large number of $100$ bond disorder realizations,
for lattice sizes $L=20-100$ for all crystal fields and disorder
strengths used in this paper, with the exception of the case
$\Delta=1.5$ and $r=0.5/1.5$, where $500$ disorder realizations
were simulated for lattice sizes $L=20-140$. For reference and
contrast, the pure-system's properties were also obtained by the
same implementation, simulating in each case at least $30$
independent runs. We close this outline of our numerical scheme
with some comments concerning statistical errors and disorder
averaging. Even for the larger lattice size studied here
($L=100$), and depending on the thermodynamic parameter, the
statistical errors of the WL method were found to be of reasonable
magnitude and in some cases to be of the order of the symbol
sizes, or even smaller. This was true for both the pure version
and the individual random-bond realizations. However, since the
nature of the present study is qualitative, not aiming to an
accurate exponent estimation, the WL errors will not be presented
in our figure illustrations. We also note that, for the
random-bond version mainly the averages over the disorder
realizations, denoted as $[\ldots]_{av}$, will be considered in
the text and their finite-size anomalies, denoted as
$[\ldots]_{av}^{\ast}$, will be used in our FSS attempts. Due to
very large sample-to-sample fluctuations, mean values of
individual maxima ($[\ldots^{\ast}]_{av}$) have not been used in
this study except for illustrative purposes, as in the case
$\Delta=1.5$, $r=0.5/1.5$ presented in our last Section.

\section{Phase diagrams: Pure and Random-Bond 2d Blume-Capel Models}
\label{sec:3}

\begin{figure}[htbp]
\includegraphics*[width=8 cm]{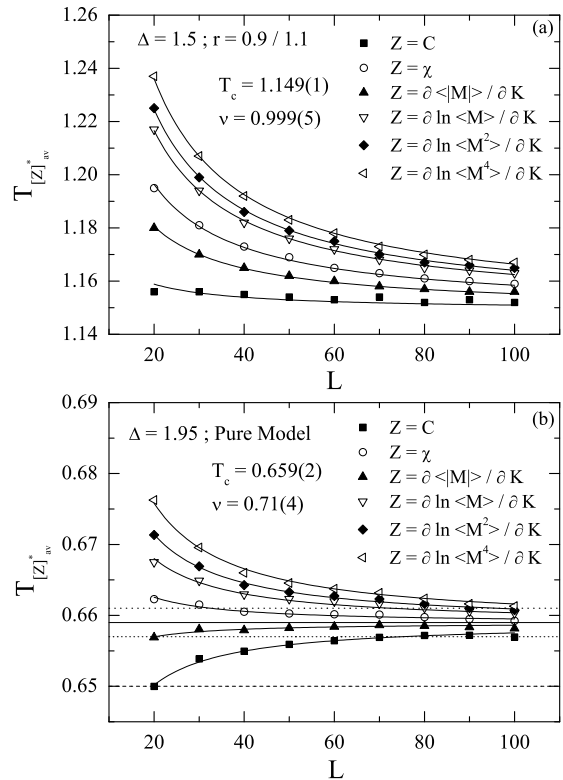}
\caption{\label{fig:1}(a) Simultaneous fitting of the six
pseudocritical temperature defined in the text for the random-bond
version for $\Delta=1.5$ and $r=0.9/1.1$. The fitting range is
$L=20-100$. (b) The same as above simultaneous fitting of the six
pseudocritical temperatures now for the pure model at
$\Delta=1.95$. The solid line shows our estimate of the critical
temperature $T_{c}=0.659$ (with the dotted lines indicating its
errors barriers) and the dashed line the estimate $T_{c}=0.650$
given in Ref.~\cite{beale86}.}
\end{figure}
This Section presents, as mentioned in the introduction, phase
diagram points for the pure and random-bond models and in the case
of the pure model compare the corresponding phase diagram points
with the existing estimates in the literature. This gives also the
opportunity to observe the reliability of our numerical approach.
Following the practice of our earlier paper~\cite{malakis09}, we
estimate phase diagram points by fitting our data to the expected
power-law shift behavior $T=T_{c}+bL^{-1/\nu}$ of several
pseudocritical temperatures. The traditionally used specific heat
and magnetic susceptibility peaks, as well as, the peaks
corresponding to the following logarithmic derivatives of the
powers $n=1,2,4$ of the order parameter with respect to the
inverse temperature $K=1/T$~\cite{ferrenberg91},
\begin{equation}
\label{eq:4} \frac{\partial \ln \langle M^{n}\rangle}{\partial
K}=\frac{\langle M^{n}H\rangle}{\langle M^{n}\rangle}-\langle
H\rangle,
\end{equation}
and the peak corresponding to the absolute order-parameter
derivative
\begin{equation}
\label{eq:5} \frac{\partial \langle |M|\rangle}{\partial
K}=\langle |M|H\rangle-\langle |M|\rangle\langle H\rangle,
\end{equation}
will be implemented for a simultaneous fitting attempt.

\begin{figure}[htbp]
\includegraphics*[width=8 cm]{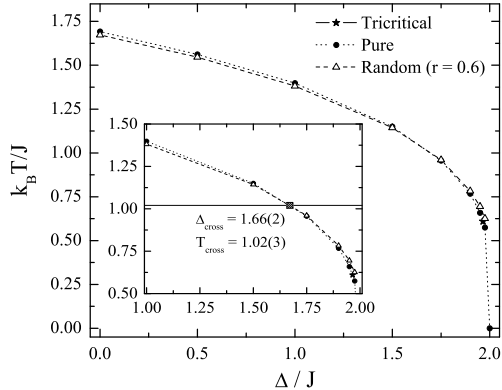}
\caption{\label{fig:2} Phase diagrams of the pure 2d BC model and
its random version at the disorder strength $r=0.75/1.25$. The
inset illustrates the crossing of the phase boundaries including
an approximate estimate of the crossing point.}
\end{figure}
Such simultaneous fitting attempts are presented in
Fig.~\ref{fig:1}. In particular Fig.~\ref{fig:1}(a) presents the
shift behavior for the random-bond 2d BC model in the weak
disorder case $r=0.9/1.1$ at the value $\Delta=1.5$ and
Fig.~\ref{fig:1}(b) illustrates the simultaneous fitting attempted
for the pure model at $\Delta=1.95$, further discussed at the end
of Sec.~\ref{sec:4b}. As noted already, the data fitted for the
random version of the model are only those of the pseudocritical
temperatures of the peaks of the averaged over disorder
thermodynamic parameters, as indicated in the figure, where the
asterisk denotes the peak of the averaged over disorder parameter
$[\ldots]^{\ast}_{av}$. The alternative route of using averages of
individual sample parameters gives almost identical estimates.
However, in cases of strong lack of self-averaging, very large
sample-to-sample fluctuations may be present. For simplicity, in
all fitting attempts, the whole range $L=20-100$ has been used.
Following this practice for the pure and random version of the 2d
BC several phase diagram points were produced. Figure~\ref{fig:2}
shows the resulting phase diagrams using our phase diagram points
for the pure 2d BC model and its random-bond version for the
disorder strength $r=0.75/1.25=0.6$. The tricritical point shown
is taken from the estimate given by Beale
$(\Delta_{t},T_{t})$=$(1.9655(10),0.610(5))$~\cite{beale86}. For
the disorder strength $r=0.75/1.25$ the points of the phase
diagram were chosen with the intension to be able to approximately
locate the emergence of the enhancement of ferromagnetic order
observed in our earlier study~\cite{malakis09} in the
ex-first-order regime at $\Delta=1.975$, where we found a
considerable increase of the critical temperature by $\sim 9\%$. A
microscopic explanation of this phenomenon, based on the
preference $s_{i}=\pm 1$ states to the strong-coupling
connectivity sites has been given in Ref.~\cite{malakis09} (see
also Ref.~\cite{kaplan08}), as a microsegregation process, due to
quenched bond randomness, that evolves continuously within the
ferromagnetic and paramagnetic phases. The inset of
Fig.~\ref{fig:2} shows that phase diagram of the random version
crosses the diagram of the pure model at approximately the
crossing point $(\Delta_{cross},T_{cross})=(1.66(2),1.02(3))$,
well before the tricritical point. The approximation of this point
was obtained by interpolation using the four points of the phase
diagram in the range $\Delta=1.0-1.9$.

Table~\ref{tab:1} summarizes the phase diagram points obtained in
this paper, by the above described traditional FFS method for both
the pure and random version of the 2d BC model for disorder
strengths $r=0.9/1.1$, $r=0.75/1.25$, and $r=0.6/1.4$, together
with corresponding phase diagram points given by
Beale~\cite{beale86} for the pure model. Note that, in Silva
\emph{et al.}~\cite{silva06} one can find an analogous table for
the pure 2d BC model including the phase diagram points given by
Beale~\cite{beale86} and the points produced in their
two-parametric WL sampling~\cite{silva06}. It can be seen that
there is an excellent coincidence of our points with those of
Beale~\cite{beale86}. The points of Beale~\cite{beale86} are based
on the very accurate phenomenological FSS scheme using a strip
geometry~\cite{nightingale82}, whereas our points are obtained via
the present simultaneous FSS analysis, based on lattices with
linear sizes $L=20-100$. The points of Silva \emph{et
al.}~\cite{silva06} are based in much smaller lattices of the
two-parameter WL sampling (linear sizes $L\leq 16$). However, in
the case $\Delta=1.95$ our phase diagram point does not agree,
within errors, with that of Beale and for this reason our
estimation with a rather generous error bars (shown also on the
panel) has been illustrated in Fig.~\ref{fig:1}(b). The rest of
Table~\ref{tab:1} contains our estimates for the random-bond
versions. One may note from this table that for small values of
$\Delta$ for instance $\Delta=1.5$ the corresponding critical
temperatures decrease as the disorder becomes stronger (compare
the three cases of disorder: $r=0.9/1.1$, $r=0.75/1.25$, and
$r=0.6/1.4$ at this value of the crystal field). On the other
hand, for $\Delta=1.9$ the trend is reversed as can be seen by
comparing the corresponding three cases of disorder. Apparently
this is a kind of reflection of the phenomenon of the enhancement
of ferromagnetic order which appears to influence the geometry of
the critical surface for the 2d random-bond BC model.

\begin{table*}
\caption{\label{tab:1}Transition temperatures of the pure and
random-bond 2d BC model obtained in this paper. Second column from
reference~\cite{beale86}, third and last entries of third and
fifth columns from Ref.~\cite{malakis09}.}
\begin{ruledtabular}
\begin{tabular}{lcccccc}
$\Delta/J$ & & & $k_{B}T/J$ &\\
\hline
& \;\;\;\;\;\;\;\;\;\;\;\;\;\;\;\;\;\;\;\;\;\;\;\;\;\;\;\;\;\;\;\;\;\;Pure & & & Random\\
& Ref.~\cite{beale86} & & $r=0.9/1.1$ &$ r=0.75/1.25$ &$r=0.6/1.4$ &\\
\hline
 0 & 1.695 & 1.693(3) & & 1.674(2) &  & \\
 0.5 & 1.567 & 1.564(3)&  & 1.547(2) &  & \\
 1 & 1.398 & 1.398(2)& & 1.381(1) &  & \\
 1.2 & & &  & & 1.277(3) & \\
 1.4 & & &  & & 1.184(3) & \\
 1.5 & 1.150 & 1.151(1) & 1.149(1) & 1.144(2) & 1.131(2) & \\
 1.6 & & & 1.084(1) & & 1.071(3) & \\
 1.7 & & & 1.005(1) &  & \\
 1.75 & & 0.958(1) & & 0.960(2) &  & \\
 1.8 & & & 0.908(1) & & 0.917(3) & \\
 1.9 & & 0.769(1) & 0.774(2) & 0.786(4) &  & \\
 1.95 & 0.650 & 0.659(2) &  & 0.702(3) &  & \\
 1.975 & & 0.574(2) & & 0.626(2) &  & \\
\end{tabular}
\end{ruledtabular}
\end{table*}

\section{Phase Transitions of the Pure and Random-Bond 2d Blume-Capel Models}
\label{sec:4}

\subsection{Strong violation of universality: The ex-first-order regime}
\label{sec:4a}

As pointed out in the introduction, in our recent investigation of
the 2d random-bond BC model~\cite{malakis09} we found dramatically
different critical behaviors of the second-order phase transitions
emerging from the first- and second-order regimes of the pure
model. Namely, different sets of critical exponents on two
segments of the same critical line appeared to describe the two
regimes: the still-second-order and ex-first-order regimes. The
study in Ref.~\cite{malakis09} was carried out for two values of
the crystal-field coupling corresponding to the second-order
($\Delta=1$) and first-order ($\Delta=1.975$) regimes of the pure
model and for the random version the disorder strength
$r=0.75/1.25$ was chosen in both cases. The strong violation of
universality observed appeared to be the result of the softening
of the first-order transition due to bond-randomness.
Specifically, it was concluded that the new strong-disorder
universality class is well described by a correlation length
exponent in the range $\nu=1.30(6)-1.35(5)$, and exponent ratios
$\gamma/\nu$ and $\beta/\nu$ very close to the Ising values $1.75$
and $0.125$, respectively~\cite{malakis09}. The above weak
universality~\cite{kim94,kim96,suzuki74} seems to be valid between
the Ising-like continuous transitions of the random-bond 2d BC
model for small values of $\Delta$ ($\Delta=1.0$ in
Ref.~\cite{malakis09}) and continuous transitions belonging to the
strong-disorder universality class.

\begin{figure}[htbp]
\includegraphics*[width=8 cm]{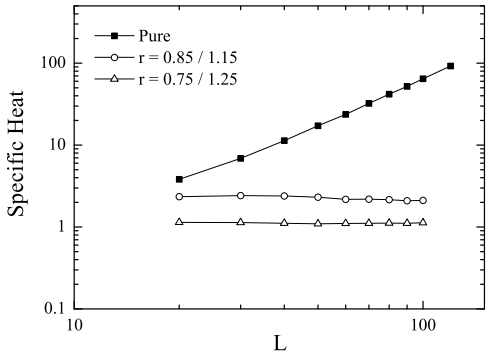}
\caption{\label{fig:3} Behavior of the random-bond 2d BC model at
$\Delta=1.975$ from Ref.~\cite{malakis09}. Illustration of the
divergence of the specific heat of the pure model (first-order
regime) and the clear saturation of the specific heat for the
random-bond (open symbols) 2d BC model for two disorder strengths
in a log-log scale.}
\end{figure}

Therefore, the strong-disorder universality class may be
characterized by the above distinct value of the correlation
length exponent and a strong saturation of the specific heat.
Qualitatively this saturating behavior is quite instructive and
for illustrative reasons is reproduced here in Fig.~\ref{fig:3}.
This figure contrasts, at $\Delta=1.975$, the specific heat's
finite-size behavior of the pure 2d BC model (first-order regime)
and two disordered cases corresponding to disorder strengths
$r=0.85/1.15$ and $r=0.75/1.25$. The saturation of the specific
heat is very clear in both cases of the disorder strength.

It is of interest to point out here that these findings for the
strong-disorder universality class appear to be fully compatible
with the classification of phase transitions in disordered systems
proposed recently by Wu~\cite{xtwu09}. According to this
classification the strong-disorder transition is expected to be
inhomogeneous and percolative with an expected exponent of the
order $\nu=1.34$~\cite{essa80}. Furthermore, it has been suggested
to us by Wu~\cite{xtwu09a}, that the strong lack of self-averaging
of this transition stems from the above properties. This violation
of self-averaging, together with the strong finite-size effects,
make the systematic MC approach of the strong-disorder regime very
demanding, if not impractical. On the other hand, the weak regime
(or Ising universality regime) suffers a much weaker lack of
self-averaging, by at least a factor of $\sim
12$~\cite{malakis09}, and a smooth behavior is observed at
moderate lattice sizes. Thus, aiming here to observe, even
approximately, the extent of the involved universality classes we
carried out our study at moderate values of the crystal field and
disorder, and found a behavior quite convincing from which the
frontier of the strong universality class can be estimated by
observing the disappearance of the expected 2d random Ising
universality class.

\subsection{Pure and random-bond 2d BC model: Range of universality with the 2d Ising model}
\label{sec:4b}

\begin{figure}[htbp]
\includegraphics*[width=8 cm]{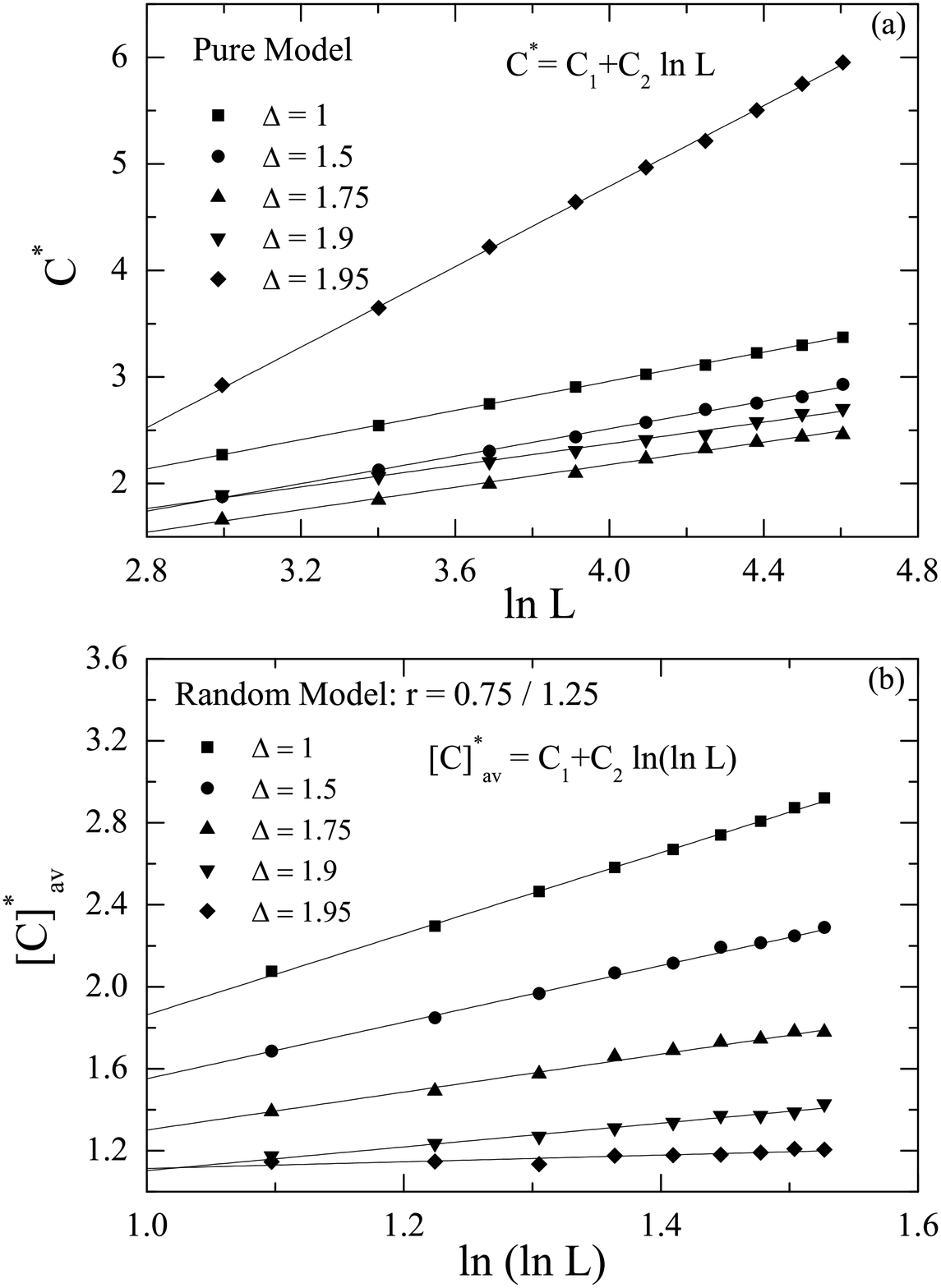}
\caption{\label{fig:4} FSS of the specific heat maxima for the
pure and random-bond 2d BC model at the same values of $\Delta$.
(a) Pure model: illustration of linear fittings assuming the
expected Ising logarithmic divergence. (b) Random model:
illustration of linear fittings assuming a double logarithmic
divergence. Note the steady fall of the double logarithmic
amplitude $C_2$.}
\end{figure}

Let us now proceed with the analysis of our numerical data for the
disorder strengths and crystal fields given in Table~\ref{tab:1}
and observe and contrast their FSS behavior with that of the pure
model. Starting this comparative study with the FSS of the
specific heat maxima (using for the random-bond version at the
strength $r=0.75/1.25$ the corresponding quantity averaged over
disorder, i.e. $[C]_{av}^{\ast}$), we present in Fig.~\ref{fig:4}
fitting attempts for the same range of $\Delta$ for the pure model
[Fig.~\ref{fig:4}(a)] and the random-bond version
[Fig.~\ref{fig:4}(b)].
As indicated by the scales in the x-axis and the functions in the
corresponding panels, the expected Ising logarithmic divergence
has been assumed for the pure model $C^{\ast}=C_{1}+C_{2}\ln L$,
whereas the double-logarithmic divergence
$[C]_{av}^{\ast}=C_{1}+C_{2}\ln(\ln L)$ has been assumed for the
random version. Although, it is very difficult to irrefutably
distinguish between a double-logarithmic divergence and a very
weak power-law divergence, the theoretically well-grounded
double-logarithmic scenario applies very well~\cite{malakis09} and
this fact can be observed also now for more values of $\Delta$ in
Fig.~\ref{fig:4}(b). There are some further features one can
observe from Fig.~\ref{fig:4}. First, for the pure model, and in
the range $\Delta>1.9$, we observe a sudden change in the behavior
of the specific heat peaks as we approach the tricritical point
which is apparently a strong cross-over effect. For the random
version and the same values of $\Delta$ no such strong effects are
noticeable and most probably the general softening effects of bond
randomness extends also to the expected cross-over phenomena
between the two different universality classes of the random 2d BC
model. From Fig.~\ref{fig:4}(b) the slopes of the
double-logarithmic fittings appear to obey a rather sensible
decreasing tendency from which we try in the next Section to
locate the frontier of strong-disorder universality class.

\begin{figure}[htbp]
\includegraphics*[width=8 cm]{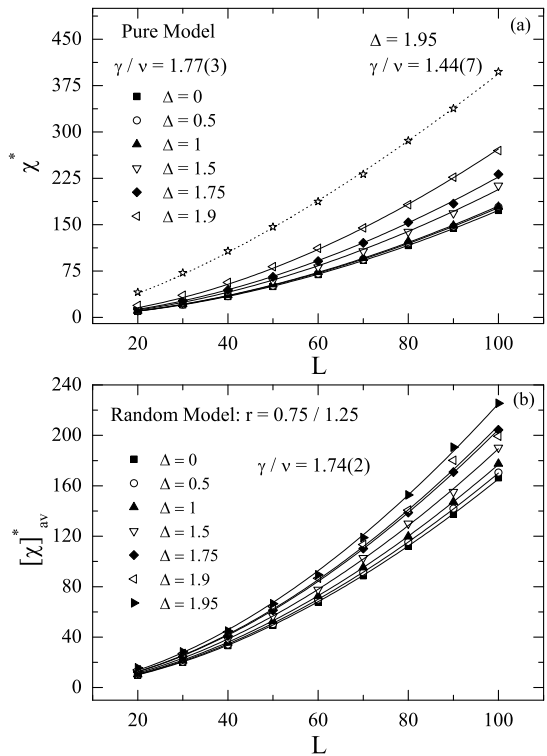}
\caption{\label{fig:5} FSS behavior of the susceptibility maxima
for the pure and random-bond 2d BC model at the same values of
$\Delta$. (a) Pure model: simultaneous fitting to a simple power
law for the first five values of $\Delta$ and a separate fitting
close to the tricritical point for $\Delta=1.95$. (b) Random
model: simultaneous fitting of the averaged susceptibility peaks
for all values of $\Delta$. Note the better behavior for the
random model and the improved estimation of the exponent ratio
$\gamma/\nu$.}
\end{figure}

A second interesting comparison follows now in Fig.~\ref{fig:5},
where again Fig.~\ref{fig:5}(a) presents the FSS of the
susceptibility maxima for the pure model, whereas
Fig.~\ref{fig:5}(b) corresponds to the random-bond version at
$r=0.75/1.25$. The influence of the exceptionally large
fluctuations in the order parameter of the pure model, as we
approach the tricritical point, is now reflected in the effective
exponent ratio $\gamma/\nu$, estimated by the simple power law
$\chi^{\ast}\sim L^{\gamma/\nu}$. These large fluctuations are a
known peculiarity of the pure model near
tricriticality~\cite{beale86}. The effective value of the exponent
is now closer to the expected value at the tricritical point
$\gamma/\nu=1.5$~\cite{beale86} than to the value of the Ising
universality class $\gamma/\nu=1.75$. For this reason a separate
power-law fitting has been applied for the value $\Delta=1.95$ in
Fig.~\ref{fig:5}(a). Note here that in Fig.~\ref{fig:5}(a) the
simultaneous fitting is applied to the first six values of
$\Delta=0 - 1.9$, whereas in Fig.~\ref{fig:5}(b) all values of
$\Delta=0 - 1.95$ are used. However, even so, the comparison of
the two panels of Fig.~\ref{fig:5}, points out the much better
limiting behavior of the random version towards the expected Ising
value $\gamma/\nu=1.75$. Therefore, we may convincingly suggest
that, for moderate values of crystal field and disorder, the Ising
universality scenario (with possible logarithmic corrections) is
well obtained in the disordered case.

\begin{figure}[htbp]
\includegraphics*[width=8 cm]{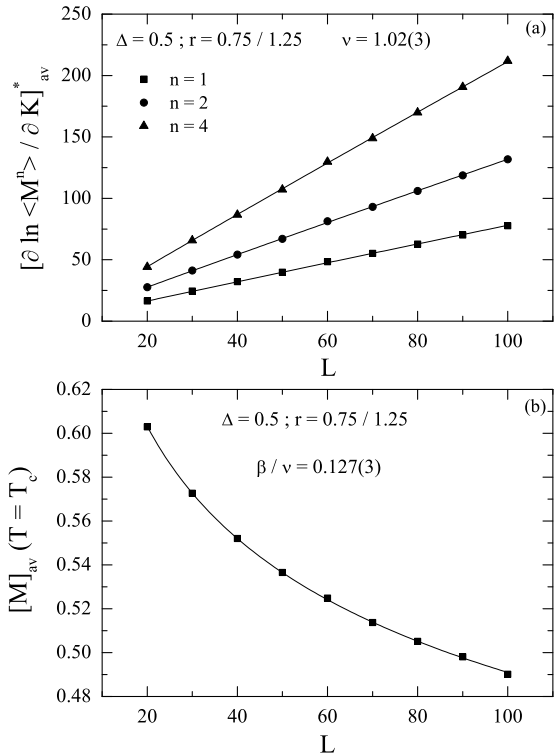}
\caption{\label{fig:6} (a) FSS analysis of the three logarithmic
derivatives ($n=1,2,4$) of the order parameter with respect to
temperature for a particular crystal field $\Delta=0.5$ and
disorder strength $r=0.75/1.25$. (b) Traditional FSS analysis of
the order parameter at the estimated critical temperature for the
same value of crystal field and disorder strength, as in panel
(a).}
\end{figure}

\begin{figure}[htbp]
\includegraphics*[width=8 cm]{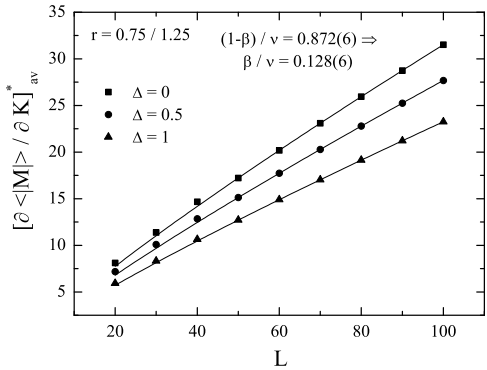}
\caption{\label{fig:7} Simultaneous fitting to a simple power law
of the averaged peaks corresponding to the absolute
order-parameter derivative for three values of $\Delta$ and
disorder strength $r=0.6$. The fitting provides an estimate of the
exponent $(1-\beta)/\nu$ and as indicated in the panel an
alternative estimate for the exponent ratio $\beta/\nu$, by
assuming $\nu=1$.}
\end{figure}

Figures~\ref{fig:6} and~\ref{fig:7} illustrate further alternative
routes, used commonly in traditional FSS analysis, that provide
clear evidence to the above suggestion namely that the
weak-disorder version belong to the 2d Ising class for suitable
moderate values of disorder and crystal-field coupling. Noteworthy
is the fact that the estimation of the critical exponents via the
traditional FSS, such as that shown in Figs.~\ref{fig:5} -
\ref{fig:7} for the random version, yields estimates very close to
the expected values, i.e. the exponents of the 2d Ising model. On
the other hand, as pointed also earlier, for the pure model, as we
approach the tricritical point ($\Delta=1.95$) the effective
exponents remind more those expected at the tricritical point than
those of the 2d Ising model.
This is particularly true for the correlation length's exponent
estimated in Fig.~\ref{fig:1}(b) by the simultaneous fitting of
the six pseudocritical temperatures. As can be seem from this
figure the estimate for the exponent $\nu$, in the case
$\Delta=1.95$, is closer to the expected tricritical value
$40/77=0.519\cdots$~\cite{beale86}, than to the 2d Ising value
$\nu=1$.

\section{Multicritical points and the strong disorder regime}
\label{sec:5}

\subsection{A novel estimation of multicritical points}
\label{sec:5a}

\begin{figure}[htbp]
\includegraphics*[width=8 cm]{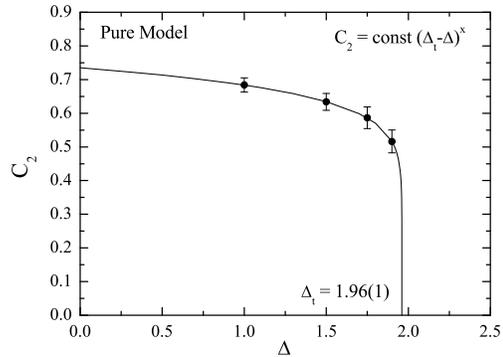}
\caption{\label{fig:8} An approximate estimation of the
tricritical value of the crystal field by fitting the decreasing
logarithmic amplitudes $C_2$ of the pure model at suitable values
of the crystal-field coupling to a suitable power law shown in the
figure.}
\end{figure}

From Fig.~\ref{fig:4}(a) one can observe the expected Ising
logarithmic divergence of the specific heat maxima. Avoiding the
value $\Delta=1.95$, which suffers from strong cross-over effects,
we attempted to estimate the tricritical value of the crystal
field by fitting the decreasing logarithmic amplitudes $C_{2}$ to
a suitable power law, as shown in Fig.~\ref{fig:8}. This may
appear a naive or questionable idea, since the behavior of
specific heat data is the Achilles' heel of FSS analysis. Yet,
Fig.~\ref{fig:8} shows that besides the large fluctuations
(errors) in logarithmic amplitudes $C_{2}$, one could
approximately estimate the tricritical crystal field
$\Delta_{t}\approx 1.96(1)$, as shown in the panel of
Fig.~\ref{fig:8}.

\begin{figure}[htbp]
\includegraphics*[width=8 cm]{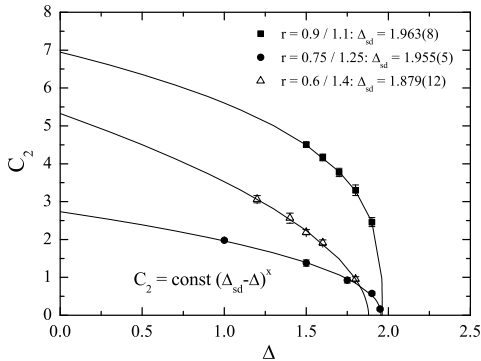}
\caption{\label{fig:9} Estimation of multicritical values of the
crystal field, where the 2d random-bond (BC) Ising universality
class meets the strong-disorder universality class. The decreasing
double logarithmic amplitudes $C_2$ of the random version has been
fitted to a power law shown and the estimates for the three
disorder strengths are shown in the panel.}
\end{figure}

From Fig.~\ref{fig:4}(b) we observe again that, the specific heat
maxima, corresponding now to the random-bond model at the disorder
strength $r=0.75/1.25$, are better-matched to the
double-logarithmic divergence than the corresponding specific heat
maxima of the pure model to a simple logarithmic divergence.
Therefore, it appears realistic to try to obtain the multicritical
point (more precisely the value of the crystal field $\Delta_{sd}$
where we expect the emergence of strong-disorder regime at a given
disorder strength), where the 2d random-bond (BC) Ising
universality class meets the strong-disorder universality class,
by fitting the decreasing double-logarithmic amplitudes to a
similar power law.
Figure~\ref{fig:9} presents now these fittings and also the
estimated multicritical values of the crystal field $\Delta_{sd}$
for the three disorder strengths considered in this paper. The
statistical errors of the corresponding double logarithmic
amplitudes $C_{2}$ are seen to be quite smaller, compared to the
corresponding amplitudes of the pure model, and the estimated
values of the values of the multicritical field, shown in the
panel of Fig.~\ref{fig:9}, appear more convincing. From their
general trend we observe that as we increase the disorder strength
the frontier of the strong-disorder universality class moves to
lower values of the crystal field, namely: ($r=0.9/1.1$,
$\Delta_{sd}=1.963(8)$), ($r=0.75/1.25$, $\Delta_{sd}=1.955(5)$),
and ($r=0.6/1.4$, $\Delta_{sd}=1.879(12)$). This behavior seems
sensible and in our opinion reflects the competition between the
ferromagnetic random interactions with the crystal field, giving a
kind of destabilization of the usual Ising-like ferromagnetic
order. A similar ground-state reflection of this competition is a
ground-state structure of unsaturated ferromagnetic ground states
discussed in our recent paper~\cite{malakis09a}. Conversely, as
disorder strength is decreased, $\Delta_{sd}$ approaches
$\Delta_{t}$, as expected.

\subsection{Strong disorder regime: The case $\Delta=1.5$, $r=0.5/1.5$ and general observations}
\label{sec:5b}

As pointed out earlier, the strong lack of self-averaging,
together with possible strong finite-size effects, make the MC
approach to the strong-disorder regime a very difficult task. The
self-averaging properties along the two segments (ex-first-order
and still-second-order) of the critical line were observed and
discussed in Ref.~\cite{malakis09}. It was shown in this paper
that the usual finite-size measure~\cite{aharony96} of relative
variance $R_{X}=V_{X}/[X]_{av}^{2}$, where
$V_{X}=[X^{2}]_{av}-[X]_{av}^{2}$ (and $X=\chi^{\ast}$ is the
susceptibility maxima), exhibits lack of self-averaging in both
cases of marginal second-order transition and the transition in
the strong-disorder or ex-first-order regime. In particular, it
was shown~\cite{malakis09} that the case studied ($\Delta=1.975$)
of the ex-first-order segment gives a much larger effect when
compared with the still-second-order regime at $\Delta=1$ and the
same disorder strength $r=0.75/1.25$ by a factor of $\sim 12$.
Therefore, phase diagram points very close to the frontier of the
strong-disorder regime may be the worst cases to study, because,
besides the very strong lack of self-averaging one may also expect
large finite-size and cross-over effects. However, such cases are
elucidatory not only for observing the intrinsic difficulties but
also for giving us the opportunity to reflect on possible links
with basic properties of the system as for instance the
ground-state structure.

\begin{figure}[htbp]
\includegraphics*[width=8 cm]{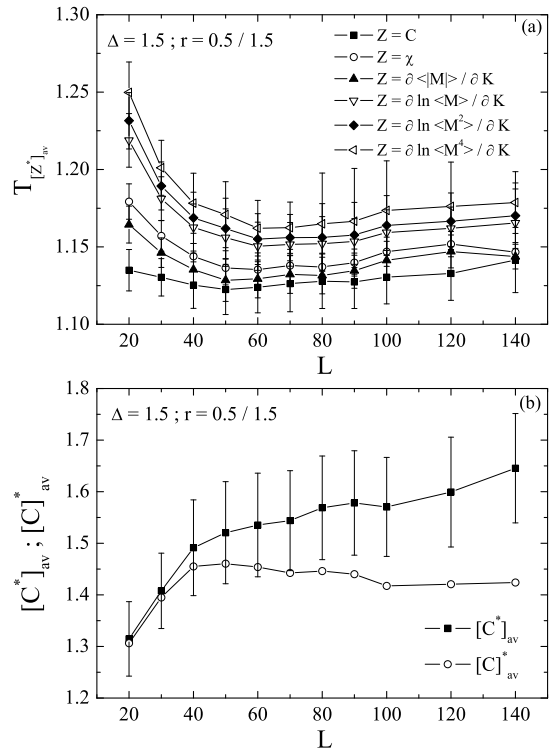}
\caption{\label{fig:10} Finite-size behavior for the case
$\Delta=1.5$ and $r=0.5/1.5$. (a) Behavior of the averaged
pseudocritical temperatures, corresponding to individual maxima,
with an illustration of the huge sample-to-sample fluctuations.
(b) Behavior of the maxima of the averaged specific heat curves
($[C]_{av}^{\ast}$) and the average of individual maxima
($[C^{\ast}]_{av}$) with their large sample-to-sample
fluctuations.}
\end{figure}

Thus, Fig.~\ref{fig:10} illustrates two important characteristics
of the case $\Delta=1.5$ and $r=0.5/1.5$. In particular,
Fig.~\ref{fig:10}(a) shows the huge sample-to-sample fluctuations
in all the pseudocritical temperatures used in this paper. The
simulation here was extended to larger lattices ($L=20-140$) and
the number of realizations studied was $500$, to be compared with
$100$ realizations studied in previous Sections (weak-regime).
Besides the enormous fluctuations, the number of $500$
realizations looks quite insufficient and strong finite-size
effects, reflected as expected in the shift behavior of the
system, produce a completely unsettled behavior.
Figure~\ref{fig:10}(b) illustrates an unusual deviating behavior
between the finite-size behavior of the maxima of the averaged
specific heat curves ($[C]_{av}^{\ast}$) and the finite-size
behavior of the average of individual maxima ($[C^{\ast}]_{av}$),
which as shown suffer large sample-to-sample fluctuations. The
behavior here resembles in many aspects the well known and still
challenging specific heat behavior of the 3d random-field Ising
model~\cite{malakis06}. However, a very clear tendency of
$[C]_{av}^{\ast}$ for a saturating behavior is observed in
Fig.~\ref{fig:10}(b) and we may speculate that this saturating
behavior is the correct asymptotic behavior for both maxima shown
in Fig.~\ref{fig:10}(b), although the behavior of
$[C^{\ast}]_{av}$ will settle down only in very large lattice
sizes, when the influence of the strong finite-size effects on the
individual maxima will diminish.

\begin{figure}[htbp]
\includegraphics*[width=8 cm]{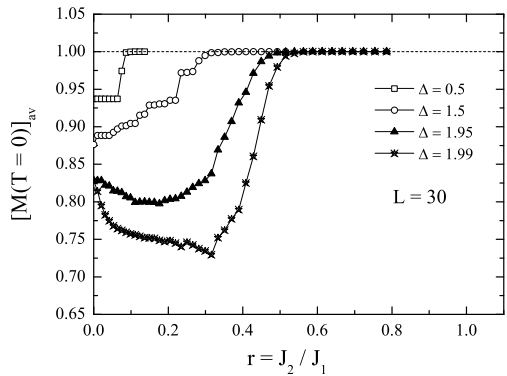}
\caption{\label{fig:11} Ground-state behavior of the order
parameter of the 2d random-bond BC model versus $r$ for various
values of $\Delta$ averaged over $250$ disorder realizations.}
\end{figure}

The above illustrations open the possibility that the case
$\Delta=1.5$ and $r=0.5/1.5$ is very close to the frontier of the
strong-universality class. This is in accordance with the trend
observed in the previous Section and can be further supported by
reproducing here some aspects of the ground-state structure of the
2d random-bond BC model. From Fig.~\ref{fig:11}, reproduced here
from Ref.~\cite{malakis09a}, one can see that approximately at
this point ($\Delta=1.5,r=0.5/1.5$), the system departs from the
ferromagnetic ground state and an unsaturated ground state is
produced, which is further enhanced with vacant sites ($s_{i}=0$)
as we increase the disorder strength. In the presence of bond
randomness the competition between the ferromagnetic interactions
with the crystal field results in a destabilization of the
ferromagnetic ground state. Depending on the realization, weak
clusters exist in $T=0$ and their points are frozen in the
$s_{i}=0$ state. This is an interesting subject, which is
presently under further consideration in both 2d and 3d by the
present authors. The novel behavior illustrated in
Fig.~\ref{fig:10} appears now as a consequence of the onset of the
unsaturated ground state at $(\Delta=1.5,r=0.5/1.5)$ which is thus
related with the critical behavior of the 2d random-bond BC model.
The presented calculation of the ground states has be carried out
in polynomially bounded computing time by mapping the system into
a network and searching for a minimum cut by using a maximum flow
algorithm (see for instance Ref.~\cite{hartmann02}) and can be
easily extended to large lattices and also to the 3d BC
random-bond model.

\section{Conclusions}
\label{sec:6}

By carrying out an extensive two-stage Wang-Landau entropic
sampling of both the pure and the random-bond 2d Blume-Capel model
we have produced phase diagram points for several disorder
strengths. Also for the pure model we found an excellent
coincidence of our points with those of Beale~\cite{beale86}. For
a particular disorder strength ($r=0.75/1.25$) we found that, as a
result of the enhancement of ferromagnetic order, the phase
diagram of the random version crosses that of the pure model at
approximately the point
$(\Delta_{cross},T_{cross})=(1.66(2),1.02(3))$, well before the
tricritical point.

The critical properties of the pure model were compared and
contrasted to those of the random model and for moderate values of
the crystal field and disorder we found that the Ising
universality scenario (with possible logarithmic corrections) is
well obtained in the case of the random version. Furthermore,
accepting in this range of couplings the assumption of the double
logarithmic scenario for the specific heat we estimated
multicritical points, where the 2d random-bond (BC) Ising
universality class meets the strong-disorder universality class.

The behavior of the strong disorder regime was also critically
discussed. The case $\Delta=1.5$ and $r=0.5/1.5$ was extensively
studied and based on the observed behavior and on the ground-state
observations we suggested that this case, most likely, lies on the
frontier between the weak- and strong-disorder universality
classes, suffering exceptionally strong finite-size effects and a
strong lack of self-averaging.

In conclusion, the present paper investigated some difficult and
important aspects of the random-bond 2d Blume-Capel model. It has
been pointed out that the behavior of this system is very
interesting and in particular the strong-disorder regime may
include many further challenges and open problems that, at the
moment, are not fully understood. In our opinion, this is a rather
complex subject deserving further research.

\begin{acknowledgments}
The authors acknowledge useful e-mail correspondence with X.T. Wu.
This research was supported by the Special Account for Research
Grants of the University of Athens under Grant No. 70/4/4071.
\end{acknowledgments}

{}

\end{document}